\documentclass[twocolumn,prl,aps]{revtex4}

\def\textbf#1{{\bf #1}}
\def\be{\begin{equation}}
\def\ee{\end{equation}}
\def\ben{\begin{eqnarray}}
\def\een{\end{eqnarray}}
\def\eea{\end{array}}
\def\bea{\begin{array}}
\newcommand{\bei}{\begin{itemize}}
\newcommand{\eei}{\end{itemize}}
\newtheorem{prop}{Proposition}
\newtheorem{obs}{Observation}

\def\ep{{\epsilon}}

\def\>{\rangle}
\def\<{\langle}

\def\ot{\otimes}

\def\rhoccq{\rho^{(ccq)}}
\def\rhoccqabe{\rho^{(ccq)}_{ABE}}
\def\rhoaabb{\rho_{AA'BB'}}
\def\sigmaccqabe{\sigma^{(ccq)}_{ABE}}
\def\sigmaccq{\sigma^{(ccq)}}
\def\sigmaab{\sigma_{AB}}
\def\sigmaabtwirl{\sigma_{AB}^{twirl}}
\def\sigmae{\sigma_E}

\begin{document}

\title{Low dimensional bound entanglement with one-way distillable 
cryptographic key}

\author{ Karol Horodecki$^{(1)}$, \L{}ukasz Pankowski$^{(2)}$, Micha\l{} Horodecki$^{(3)}$, Pawe\l{} Horodecki$^{(4)}$}

\affiliation{$^{(1)}$Department of Mathematics Physics and Computer Science, University of Gda\'nsk, 80--952 Gda\'nsk, Poland}

\affiliation{$^{(2)}$Institute of Theoretical Physics and Astrophysics University of Gda\'nsk, 80--952 Gda\'nsk, Poland}

\affiliation{$^{(3)}$Institute of Theoretical Physics and Astrophysics University of Gda\'nsk, 80--952 Gda\'nsk, Poland}

\affiliation{$^{(4)}$Faculty of Applied Physics and Mathematics,
Technical University of Gda\'nsk, 80--952 Gda\'nsk, Poland}

\begin{abstract}
We provide a class of bound entangled states  that have positive distillable secure key rate. 
The smallest state of this kind is $4 \otimes 4$, which shows that peculiar security 
contained in bound entangled states does not need  high dimensional systems. We show, that for 
these states a positive key rate can  be obtained by {\it one-way} Devetak-Winter 
protocol. Subsequently the volume of bound entangled key-distillable states in arbitrary  dimension 
is shown to be nonzero.  
We provide a scheme of verification of cryptographic quality of experimentally prepared  state in terms of local observables. Proposed set of $7$ collective settings is proven to be optimal in number of settings.
\end{abstract}

\maketitle

Quantum cryptography is one of the very interesting practical
phenomena within quantum information theory 
\cite{BouwmeesterEZ-book,Nielsen-Chuang,Alber2001,Gruska}.
There were in general two ideas to produce cryptographic key.
The first was based on sending nonorthogonal states
\cite{BB84}, the second - on specially chosen measurements
of maximally entangled pairs \cite{Ekert91}.
They have been shown to be equivalent in general
\cite{Shor-Preskill} including most general eavesdropper attack.
An important ingredient of the protocol was so called 
Quantum Privacy Amplification \cite{QPA} based on
distillation of EPR paris \cite{BBPSSW1996}. Despite of natural
expectations, that distillability of EPR pairs is a precondition 
of secure key it has been recently shown \cite{pptkey,keyhuge} 
that the class of states which contain ideal key ({\it private states}) is 
much wider than class of maximally entangled states. It has been shown
that certain {\it bound entangled} (BE) states \cite{bound}
(from which no EPR pair can be distilled) can be distilled to (approximate)
private states. In other words - surprisingly - 
there are bound entangled states with $K_D >0$. 
    
However  the BE states  with nonzero distillable key $K_D>0$ of  
Ref. \cite{pptkey,keyhuge} require  Hilbert space of quite large 
dimension, which makes impression that BE states  useful for cryptography 
are rather exceptional, and far from experimental regime. 
In this paper we provide a general construction of class of BE states with
$K_D>0$ and give examples of $4\otimes 4$ states having this property.
To this end we consider binary mixture $\{p_1,p_2\}$ 
of two orthogonal private bits (private states with at least one bit of ideal key).
We first show that any such non equal mixture has $K_D >0$, and can be
distilled by - quite remarkably - {\it one-way} Devetak-Winter protocol \cite{DevetakWinter-hash}.
Next we construct special pairs of private bits, and show that for certain probability their mixture is key distillable and bound entangled. To obtain this 
we assure that the state remains positive (and even invariant) under partial transposition (PPT) \cite{Peres96}
which is sufficient condition for a state to be non distillable  \cite{bound}. We then provide
an example of the smallest state of our construction which resides on 4 qubits. 
Basing on this family of BE key distillable states we argue, 
that the volume of BE states with $K_D>0$ is nonzero in arbitrary dimension. 
We exploit their properties, and consider their experimental preparation. 
We then show how to verify that experimentally prepared state has nonzero 
distillable key. Finally the optimal decomposition of observables needed 
for veryfication if certain $4\ot 4$ state has $K_D >0$ is given. 
 

Let us first recall that private bit \cite{keyhuge} is  a state 
$\gamma_{AA'BB'}$, where $AA'$ is hold by Alice, $BB'$ by Bob, 
and measurement of $AB$ in standard basis provides one bit of perfect key
(ie. possible outcomes are either $\{00\}$ or $\{11\}$
with equal probability $1/2$ and they are completely independent on
measurements on physical system different than $AA'BB'$).
Following \cite{keyhuge} we will write $\gamma$ in matrix representation in $ABA'B'$ order of subsystems, so that its $4\times 4$ structure corresponds to
two qubit subsystem $AB$. 

We shall start the construction of key distillable BE states by providing a class of states with $K_D>0$:

{\prop .- Consider two private bits $\gamma_{1}$, 
$\gamma_{2}$ and take any biased mixture of the form:
\begin{equation}
\rho= p_{1}\gamma_{1} + p_{2}\sigma^{A}_{x}\gamma_{2}\sigma^{A}_{x} 
\label{eq:rho-prop}
\end{equation}
with, say, $ p_{1} > p_{2} $ and 
$ \sigma^{A}_{x} = [\sigma_{x}]_{A} \otimes I_{A'BB'}$. 
The distillable key $K_D(\rho)$ fulfills $K_D(\rho) \geq 1- h(p_1)$
where $h(p_1)$ is the binary entropy of distribution $\{p_1, p_2\}$.
\label{prop:mixedpbits}
}

Before we prove the proposition, we have to recall a technique called "privacy squeezing" 
\cite{keyhuge} that allows to investigate privacy of states of type $\rho_{AA'BB'}$.
The state is purified to $\psi_{ABA'B'E}$ so that Eve holds the subsystem $E$ of 
$\psi$. To draw key, Alice and Bob will measure systems $AB$ in standard basis,
and will process the outcomes by public discussion. The systems $A'B'$ 
will not be actively used, so the relevant state is 
\be
\rho_{ABE}^{(ccq)} = \sum_{ij} p_{ij} |ij\>\<ij|_{AB} \ot \rho^{(ij)}_E.
\ee
where $\rho^{ij}_E$ are Eve's states given the outcome was $ij$. The state is called $ccq$
(two registers are classical).
From such state by Devetak-Winter \cite{DevetakWinter-hash} protocol one can get 
\be
K^{DW}=I(A:B)-I(A:E)
\label{eq:dwrate}
\ee
bits of key where $I(A:B)_{\rho} = S(A) + S(B) - S(AB)$, $S(X)$ being von Neumann 
entropy of $X$ subsystem of state $\rho$. We will now provide a different 
ccq state $\sigmaccqabe$, which is 
no better, in the sense that Eve can obtain it from $ \rhoccqabe$ 
by some operation on her system. Clearly, such $\sigmaccq$  can give no more key than 
$\rhoccq$. The state $\sigmaccq$ we produce as follows. First, from $\rhoaabb$ 
we obtain its {\it privacy squeezed} (p-squeezed) version $\sigmaab$. To this end  $(i)$ we apply so-called twisting 
\cite{pptkey,keyhuge}  i.e. a unitary  transformation controlled by standard basis on $AB$ 
$U_\tau=\sum_{ij} |ij\>\<ij|_{AB}\ot U^{ij}_{A'B'}$  and subsequently
$(ii)$ we trace over systems $A'B'$. One finds \cite{pptkey,keyhuge}
that  p-squeezed state $\sigmaab$ has a property that 
the ccq state $\sigmaccqabe$ emerging after measuring it in standard basis $|ij\>$ 
is no better than $\rhoccqabe$. However, if twisting is properly chosen,
then it may still produce much key. The privacy is now squeezed to solely two systems 
$AB$.

{\bf Proof  of Proposition.}
Using the above method, we will apply p-squeezing to the state $\rho$, and 
show that from the $\sigmaccq$ the DW protocol gives $1-h(p_1)$ rate of key. 
A basic fact about private bits \cite{pptkey} is that there exists twisting 
which brings them to the form $\psi_0^{AB}\ot \tilde \rho_{A'B'}$
where $\tilde\rho_{A'B'}$ is some state,  $\psi_0$ is one of four Bell states 
\ben
|\psi_{0,1}\> = {1\over \sqrt{2}}(|00\> \pm |11\>) \\
|\psi_{2,3}\> = {1\over \sqrt{2}}(|01\> \pm |01\>)
\een
and in the twisting $U_{01}=U_{10}=I$. 
Using this, we immediately obtain a twisting which, followed by 
partial trace over $A'B'$, turns the state $\rho$ 
of eq. (\ref{eq:rho-prop}) into a mixture of $\psi_0$ and $\psi_2$. 
 I.e. the p-squeezed state of (\ref{eq:rho-prop}) is $\sigma_{AB} = p_1 |\psi_0\>\<\psi_0| + p_2 |\psi_2\>\<\psi_2| $. 
Its purification is of the form 
\be
\psi'_{ABE}= \sqrt{p_1}|\psi_0\>_{AB}|e_1\>_{E} + \sqrt{p_2}|\psi_2\>_{AB}|e_2\>_{E}
\label{psqueezed}
\ee
so that by measuring it in  standard basis Alice and Bob obtain the ccq state
\be
\sigmaccqabe = {p_1\over 2}[P_{00} + P_{11}]\ot P_{e_1} + 
{p_2\over 2}[P_{01} + P_{10}]\ot P_{e_2}
\ee
with $P_{ij}=|ij\>\<ij|$ and $P_{e_i}=|e_i\>\<e_i|$.
For this state, we have $I(A:B)-I(A:E)= 1-h(p_1)$ Thus by DW protocol \cite{DevetakWinter-hash}
we get this amount of key.
Because this state is no better than our state of interest $\rhoccqabe$
(the ccq state obtained by measuring $AB$ systems of initial state $\rho$)
we obtain that the distillable key of $\rho$ satisfies $K_D(\rho)\geq 1-h(p_1)$, 
which ends the proof of proposition. 
 
Let us note that the key is here drawn by one-way protocol. We have found that 
applying two-way recurrence protocol will not increase the key rate. 

{\it Construction of small dimensional bound entangled states with $K_D>0$}
First let us recall, that any private bit can be represented in its $X$-form \cite{keyhuge}:
\begin{eqnarray}
\gamma_{ABA'B'}={1\over 2}\left[
\begin{array}{cccc}
\sqrt{XX^{\dagger}} &0&0& X \\
0& 0&0&0 \\
0&0&0& 0\\
X^{\dagger}&0&0& \sqrt{X^{\dagger}X}\\
\end{array}
\right]
\label{xform}
\end{eqnarray}
where $X$ is arbitrary operator with trace norm one, 
and which completely represents the pbit. Then, we obtain that state 
$\rho$ is of the form 
\begin{eqnarray}
\rho= \frac{1}{2}\left[ \begin{array}{cccc}
p_{1}\sqrt{X_{1}X_{1}^{\dagger}} & 0 & 0 & p_{1}X_{1} \\
0 & p_{2}\sqrt{X_{2}X_{2}^{\dagger}}& p_{2} X_{2} & 0 \\
0 & p_{2}X_{2}^{\dagger} & p_{2} \sqrt{X_{2}^{\dagger}X_{2}} & 0\\
p_{1}X_{1}^{\dagger} & 0 & 0 & p_{1}\sqrt{X_{1}^{\dagger} X_{1}}\\
\end{array}
\right].
\label{standform}
\end{eqnarray}

We assume now that $A'$ and $B'$ are systems both described 
by ${\cal C}^{d}$. Now the essential part of the construction 
is the following substitution: $X_{1} = {1\over ||W_U||}W_U $
where
\begin{equation}
W_U=\sum_{ij}u_{ij} |ij\rangle \langle ji| 
\end{equation}
and $u_{ij}$ are unitary matrix elements of some matrix
$U$ on ${\cal C}^{d}$. Note that $|| W_U||=\sum_{ij}|u_{ij}|$
(here we use the trace norm of the matrix).
The second operator we choose 
$X_{2} = \frac{W_U^{\Gamma}}{||W_U^{\Gamma}||}$ with
$\Gamma$ being partial transposition on subsystem $B'$. In this case
one has just $||W_U^{\Gamma}||=d$

The corresponding mixing probabilities are
\begin{eqnarray}
&& p_{1}=\frac{||W_U||}{||W_U||+||W_U^{\Gamma}||}\quad 
p_{2}=\frac{||W_U^{\Gamma}||}{||W_U||+||W_U^{\Gamma}||} 
\end{eqnarray}
respectively.

Collecting two simple observations, namely that:
(i) $[W_U W_U^{\dagger}]^{1\over 2}= 
[W_U^{\dagger}W_U]^{1\over 2}=\sum_{ij} |u_{ij}| 
|ij \rangle \langle ij|$ (which after normalisation 
by factor $||W_U||$ gives separable, PPT-invariant 
state),(ii) $[W_U W_U^{\dagger}]^{1\over 2}= 
[W_U^{\dagger}W_U]^{1\over 2}=\sum_{i} | ii \rangle \langle ii|$, 
(again after normalisation giving  PPT-invariant separable state)
We get immediately that $\rho$ with 
parameters defined like above is PPT invariant.
At the same time we have desired security condition
$p_{1}>p_{2}$ 
if only 
\begin{equation}
\frac{p_{1}}{p_{2}}=\frac{||W_U||}{||W_U^{\Gamma}||} \equiv
\frac{\sum_{ij} |u_{ij}|}{d} > 1.
\label{rate}
\end{equation}
The latter is satisfied for {\it any} unitary U which written in $\{|ij\>\}$ basis
has more than $d$ nonzero entries.

Thus we have a large class of states that contain secure 
key and are at the some time PPT invariant states.
Of course they are entangled since entanglement 
is a precondition of secure key distillation \cite{Curty04:key-ent}.
{\obs
The ratio of $p_1$ and $p_2$ in (\ref{rate}) which is related to
$K^{DW}$ key rate achieves the highest value for unimodular 
unitaries $U$ (ie such that $|u_{ij}|=\frac{1}{\sqrt{d}}$  
irrespectively of indices $\{ i, j\}$). Then it 
amounts to $[\frac{p_{1}}{p_{2}}]_{optimal}= \sqrt{d}$.
}

Indeed, by use of Lagrange multipliers with slightly more general constraints 
$\sum_{ij} |u_{ij}|^2 = d$ we get that optimal $U$ is unimodular. 


{\it Example of small bound entangled states with  $K_D>0$ on 
$4 \otimes 4$ system.}  Putting $d=2$ we get the smallest secure BE states in our 
construction. An easy example is a state with $U$ equal to 
1-qubit Hadamard gate (H). Note, that in this case, the state $\gamma_2$
coincides with the so called "flower state" with $U=H$, which exhibited locking of
entanglement cost \cite{lock-ent}.  The  total state can be written as a 
mixture of Bell states on $AB$ subsystem of state, that are classically correlated 
with some other states on $A'B'$. Namely we have
\be
\rho_H = \sum_{i}q_i |\psi_i\>\<\psi_i|_{AB}\ot \rho_{A'B'}^{(i)}
\label{eq:ex}
\ee
where the correlated states are the following: 
\ben
&&\rho^{(0)} = {1\over 2}[P_{00} + P_{\psi_2}] \nonumber  \\
&&\rho^{(1)} = {1\over 2}[P_{11} + P_{\psi_3}] \nonumber \\
&&\rho^{(2,3)} = P_{\chi_\pm}
\een
with $P_{\psi_i}$ being projectors onto corresponding Bell states and $P_{\chi_\pm}$ projectors onto pure states
\be
\chi_\pm ={1\over 2} (\sqrt{2 \pm  \sqrt{2}}\,\, |00\> \pm \sqrt{2 \mp \sqrt{2}}\,\,|11\> )
\ee
respectively. 
The mixing distribution $\{q_i\}_{i=0}^{3}$ is $\{ {p_1\over 2}, {p_1\over 2}, {p_2\over 2},{p_2\over 2} \}$. 
Since $d = 2$, one has $ p_1 = {\sqrt{2}\over 1 + \sqrt{2}} $, 
so by proposition (\ref{prop:mixedpbits}) a  
positive key rate can be gained from this $4$-qubit PPT state. It reads:
\be 
K_D^{DW}(\rho_H) = 1 - h(p_1) = 0.0213399
\ee 
per copy of $\rho_H$. Note again, that it automatically means, that
$\rho_H$ is bound entangled. Indeed we have shown, that state of that 
construction is PPT invariant, and it can not be separable, as it has 
non-zero distillable key.  We show now the next property of this state.

{\obs  The state $\rho_H$ is extremal in the set of PPT states. 
\label{obs:extreme}
}

$\rho_H$ is a mixture of form (\ref{standform}) 
with $X_1 =W_H$ and $X_2=X_1^{\Gamma}$. It is straightforward to check, that any
mixture with the weight of $\gamma_1$ different 
then value  $p_1 = {\sqrt{2}\over 1 + \sqrt{2}}$ leads to NPT state. In fact the 
same argument proves extremity  of the state from our construction with $X=W_U$, 
if only $X$ is hermitian operator, and either  $X$ or $X^{\Gamma}$ has some 
positive eigenvalue.

Although the states $\rho_U$ lay on the edge of the set of PPT states, 
basing on this family of states 
we are able to show, that the set of PPT key distillable states has non-zero
volume in the set of all PPT $4\ot 4$ states . 
{\obs The set of PPT distillable key of the form $\rho_U$ has 
nonzero volume in $4\ot 4$.
\label{obs:nonzero}
}

The proof of this observation bases on the fact, that Devetak-Winter 
lower bound (\ref{eq:dwrate}) is continuous in $\rho$,
as mutual information is continuous function of $\rho$. If we consider 
now $ p_{mix} \rho + (1-p_{mix}) {I\over 16}$,
for suitably small $p_{mix}$ there exists suitably small $\ep > 0$ 
for which the ball of all states with the center
at $\rho_U$ and radius $\ep$ lays within the set of PPT states. 
By continuity argument, one gets the thesis.
In fact a similar argument gives this result in any
dimension higher than $4\ot4$.

{\it  Upper  bound for $K_D$.-} Similarly as private bit is 
analogue of singlet from entanglement distillation theory, 
the state $\rho$ of (\ref{eq:rho-prop}) is analogue of mixture of two singlets 
(which is actually just the p-squeezed state $\sigma$).
The distillable entanglement of the latter state  is just $1-h(p_1)$, \cite{BDSW1996} and 
moreover it is achieved, by coherent application of the DW protocol. 
Therefore there is a question, if $K_D(\rho)$ is just equal to $1-h(p_1)$.
In \cite{pptkey,keyhuge} we show that relative entropy of entanglement $E_r$ 
is an upper bound for $K_D$. We have analysed the $4\ot 4$ 
PPT state $\rho_H$, and have found a separable state which 
gives $E_r(\rho_H)\leq 0.116$.

{\it Preparation of $\rho_U$  $4\ot 4$ states.-}
As it is shown in example (\ref{eq:ex}) $\rho_U$ $4\ot 4$ states  
are only classically correlated along
$AB$ versus $A'B'$ cut. Thus they can be created by 
preparing randomly according to $\{q_i\}$ distribution, 
separately two states: a Bell state $\psi_i$ (for $AB$ subsystem) 
and $\rho^{(i)}$ (for $A'B'$ subsystem). 
According to observation(\ref{obs:extreme}) $\rho_U$ lays  
on the boundary of PPT set, any small perturbation can
destroy this property. However in spirit
of observation (\ref{obs:nonzero}), one can construct PPT 
states that are to some extent robust against perturbations. 

{\it Key distillability verification for experimentally prepared state.-}
	We now address the question of verification whether a state
 prepared experimentally in many copies has nonzero distillable key. 
 In the spirit of proof of the proposition (\ref{prop:mixedpbits}) instead of 
estimating whole $\rho_{ABA'B'}$  we suggest to estimate 
only few parameters of its privacy squeezed state $\sigmaab$ and subsequently 
compute some lower bound on the value of  DW rate (\ref{eq:dwrate}) for 
the ccq state $\sigmaccqabe$ of the latter. 

Once we do not estimate whole state $\rho_{ABA'B'}$, the formula 
 (\ref{eq:dwrate}) can not be used directly to decide its quality. 
Instead we first consider a lower bound for distillable key from 
$ccq$ state. Namely
\be
K_D(\rho^{(ccq)}) \geq I(A:B)_{\rho^{(ccq)}} - S(E)_{\rho^{(ccq)}},
\ee
which is a consequence of formula (\ref{eq:dwrate}).  
Indeed, $I(A:E)$ for ccq state is 
equal to $S(\rho_E) - \sum_i p_i S(\rho_E^i)$, which is 
Holevo function of ansamble $\{p_i,\rho_E^i\}$. This however can not 
be greater then just entropy of Eve's subsystem $S(\rho_E)$ 
and the assertion follows.

Using this lower bound we provide now the one which is a function of 
only diagonal and antidiagonal matrix elements 
of (2-qubit) privacy squeezed state $\sigmaab$.
Note, that $S(\sigmae) = S(\sigmaab)$, as the total $ABE$ state is pure.
To estimate $S(\sigma_E)$ we will consider state $\sigmaab$ subjected to 
twirling \cite{BBPSSW1996}  which projects onto Bell basis. Twirling can not 
decrease the entropy, and commutes with  measurement in computational basis, so one has that 
\be
K_D(\sigmaccqabe) \geq I(A:B) - S(\sigmaabtwirl)
\label{eq:llowerbound}
\ee
This is desired lower bound as it
is a function of only those parameters, which we suggest to  
estimate experimentally.

Although the formula (\ref{eq:llowerbound}) is useful for
one-way key distillable states, knowing the diagonal and antidiagonal
elements of $\sigma_{AB}$ is enough to decide if a two-way recurrence 
protocol can make key rate non-zero. 
 
We give now the observables, which measured on $\rho_{ABA'B'}$ reveals desired 
elements of its p-squeezed state. These are:
\ben
&&O_1 = U_\tau^{\dagger} [\sigma_z \ot \sigma_z]_{AB}\ot I_{A'B'} U_\tau \nonumber\\
&&R_{1,2} = U_\tau^{\dagger} [P_{\psi_{0,2}} - P_{\psi_{1,3}}]_{AB}\ot I_{A'B'} U_\tau \nonumber\\
&&I_{1,2}  = U_\tau^{\dagger} [\tilde{P}_{\psi_{0,2}} - \tilde{P}_{\psi_{1,3}}]_{AB}\ot I_{A'B'} U_\tau \nonumber\\
\een
where $\tilde{P}_{\psi_k}$ are Bell states with relative phase $\pm i$. The observable $O_1$ reveals 
the diagonal elements of state $\sigma_{AB}$. In fact, it is just equal to $[\sigma_z \ot \sigma_z]_{AB}\ot I_{A'B'}$
as any twisting (here $U_\tau^{\dagger}$) commutes with the measurement 
in basis which it controls \cite{keyhuge}. $O_1$ needs therefore
just one setting. The $R_k$ ($I_k$) observables reveals real (imaginary) parts 
of (possibly complex) coherences on antidiagonal of $\sigma_{AB}$. 
Indeed, one has for example
\ben
&&Tr R_1 \rho_{ABA'B'} = \nonumber \\
&&Tr U_\tau^{\dagger} [P_{\psi_0} - P_{\psi_1}]_{AB}\ot I_{A'B'} U_\tau \rho_{ABA'B'}=\nonumber \\
&&Tr  [P_{\psi_0} - P_{\psi_1}]_{AB}\ot I_{A'B'} U_\tau \rho_{ABA'B'}U_\tau^{\dagger}=\nonumber \\
&&Tr  [P_{\psi_0} - P_{\psi_1}] Tr_{A'B'}[U_\tau \rho_{ABA'B'}U_\tau^{\dagger}]=\nonumber \\
&&Tr [P_{\psi_0} - P_{\psi_1}] \sigma_{AB}.
\een
where second equality is by property of trace, and third by 
definition of subsystem of a quantum state. 
Last equality uses definition of privacy squeezed state ($\sigma_{AB}$) 
which is obtained 
by acting on $\rho_{ABA'B'}$ with some twisting $U_\tau$ 
and tracing out $A'B'$ subsystem.

{\it Local decomposition of verification observables for $\rho_H$.}
In case of $\rho_H$ twisting which realises privacy squeezing is equal to:
\be
U_\tau = |00\>\<00|\ot W_H + |01\>\<01| \ot P_H + (|10\>\<10| + |11\>\<11|)\ot I
\ee
where by $P_H = \sum_{ij=0}^1 h_{ij}|ii\>\<jj| + \sum_{i\neq j} |ij\>\<ij|$ for $h_{ij}$ 
elements of Hadamard and $W_H = P_H^{\Gamma}$. 
For this $U_\tau$,
the observables $R_1$ and $I_1$ can be decomposed into Pauli operators $\{I,\sigma_x ,\sigma_y, \sigma_z\}$ in the 
following way:
\ben
&R_1& = {1\over 4}[\sigma_x \ot \sigma_x -\sigma_y \ot \sigma_y]_{AB} \ot \nonumber \\ 
&&\ot\, [(I\ot \sigma_z + \sigma_z\ot I) + (\sigma_x \ot \sigma_x + \sigma_y\ot \sigma_y)]_{A'B'} 
\nonumber \\ 
&I_1& = -{1\over 4}[\sigma_x \ot \sigma_y + \sigma_y \ot \sigma_x]_{AB}\ot  \nonumber \\
&&\ot\, [(I\ot \sigma_z + \sigma_z\ot I) + (\sigma_x \ot \sigma_x +\sigma_y \ot \sigma_y)]_{A'B'}
\nonumber 
\een
and the same for $R_2$ and $I_2$, which read
\ben
&&R_2 = {1\over 4}[\sigma_x \ot \sigma_x +\sigma_y \ot \sigma_y]_{AB}  \ot [(I\ot I - \sigma_z\ot\sigma_z)+  \nonumber \\
&&{1\over \sqrt{2}}((I\ot \sigma_z + \sigma_z\ot I)  + (\sigma_x \ot \sigma_x +\sigma_y \ot \sigma_y))]_{A'B'}\nonumber \\
&&I_2 = -{1\over 4}[\sigma_y \ot \sigma_x - \sigma_x \ot \sigma_y]_{AB}
\ot [(I\ot I - \sigma_z\ot\sigma_z)+  \nonumber \\
&&{1\over \sqrt{2}}((I\ot \sigma_z + \sigma_z\ot I) + (\sigma_x \ot \sigma_x +\sigma_y \ot \sigma_y))]_{A'B'}\nonumber
\een

Generalizing approach for two-qubit case of \cite{opt-local-decomp,opt-multi-local-decomp} to four qubit case, one can easily show, 
that decomposition for $R_2$ and $I_2$ is optimal in the sense, 
that it needs $6$ settings. The set of these 6 collective settings 
is enough for both  $R_2, I_2$ and $R_1, I_1$ (for the latter it needs 
only different classical post-processing). Since this set of settings is optimal for 
determining $R_2,I_2$ and suffices for determining all $R_i, I_i$, we conclude, that 
it is optimal for our task. Together with one setting for $O_1$ one needs 7 different
collective settings to verify via lower bound (\ref{eq:llowerbound}) if the state $\rho_H$ has
non zero distillable key. 
\vspace{1cm}

We thank Ryszard Horodecki for helpful comments and remarks.
KH, MH and PH acknowledge the hospitality of 
Isaac Newton Institute for Mathematical Sciences during the QIS programme. 
The work is supported by Polish Ministry of Scientific Research and Information
Technology under the (solicited) grant no.~PBZ-MIN-008/P03/2003 and by
EC grants RESQ, contract no.~IST-2001-37559 and QUPRODIS, contract
no.~IST-2001-38877.
\bibliographystyle{apsrev}
\bibliography{refmich}

\end{document}